\documentstyle[psfig]{l-aa}     
\def\ref{\par \noindent \hang}

\def\etal{et al.\thinspace}
\def\eg{{\it e.g.\ }}

\def\ie{{\it i.e.\ }}
\def\approxlt{\mathrel{\hbox{ \lower .5ex \hbox {$\sim$}
	\llap{\raise .15 ex \hbox{$<$}} }}}
\def\approxgt{\mathrel{\hbox{ \lower .5ex \hbox {$\sim$}
	\llap{\raise .15 ex \hbox{$>$}} }}}

\def\multleft#1{\hbox to size{\vbox {\halign {\lft{##}\cr #1}}\hfill}\par}
\def\multright#1{\hbox to size{\vbox {\halign {\rt{##}\cr #1}}\hfill}\par}

\def\degmark{$^\circ$}
\def\today{\ifcase\month\or January\or February\or March\or April\or May\or
      June\or July\or August\or September\or October\or November\or December\fi
      \space\number\day, \number\year}
\def\<{\thinspace}
\def\s{\hbox{\phantom{5}}}      

\def\boxit#1{\vbox{\hrule\hbox{\vrule\kern3pt\vbox{\kern3pt
	  #1 \kern3pt}\kern3pt\vrule}\hrule}}

\def\km{{\rm\thinspace km}}

\def\Mpc{{\rm\thinspace Mpc}}

\def\s{{\rm\thinspace s}}

\def\kmps{\hbox{$\km\s^{-1}\,$}}

\def\kmpspMpc{\hbox{$\kmps\Mpc^{-1}$}}

\def\H2{\hbox{H$_{2}$~}}

\begin{document}

\thesaurus{03(11.02.1; 11.02.2; 11.14.1; 11.17.3; 13.09.1)}

\title{Near--infrared imaging of the host galaxies of flat spectrum radio 
quasars\thanks{Based on observations collected at the European Southern 
Observatory, La Silla, Chile.}}

\author{Jari K. Kotilainen \inst{1,2}, Renato Falomo\inst{3} and Riccardo 
Scarpa\inst{4}}
\offprints{J.K. Kotilainen (SISSA address)}

\institute{International School for Advanced Studies (SISSA), via Beirut 
2--4, I--34014 Trieste, Italy; e-mail: jkotilai@sissa.it
\and
Tuorla Observatory, University of Turku, V\"{a}is\"{a}l\"{a}ntie 20, 
FIN--21500 Piikki\"{o}, Finland
\and
Osservatorio Astronomico di Padova, vicolo dell'Osservatorio 5, I--35122 
Padova, Italy; e-mail: falomo@astrpd.pd.astro.it
\and
Space Telescope Science Institute, 3700 San Martin Drive, Baltimore, MD 21218,
U.S.A; e-mail: scarpa@stsci.edu
}

\date{Accepted 24 October 1997; received 23 June 1997}

\maketitle

\markboth{J.K. Kotilainen, R. Falomo \& R. Scarpa: NIR imaging of FSRQ hosts}{}

\begin{abstract}

We present the results of a high resolution (0.27$''$ px$^{-1}$) 
near--infrared H band (1.65 $\mu$m) imaging survey of a complete sample of 20 
flat radio spectrum quasars (FSRQ) extracted from the 2Jy catalogue of radio 
sources (Wall \& Peacock 1985). The observed objects are intrinsically 
luminous with median M(B) = --25.5. The median redshift of the objects in the 
sample is z = 0.65. At this redshift, the H band observations probe the old 
stellar population of the hosts at rest frame wavelength of $\sim$1 $\mu$m. 

We are able to detect the host galaxy clearly for six (30 \%) FSRQs and 
marginally for six (30 \%) other FSRQs, while the object remains unresolved 
for eight (40 \%) cases. We find the galaxies hosting FSRQs to be very 
luminous (M(H)$\sim$--27). Compared with the typical galaxy luminosity L* 
(M*(H)$\sim$--25) they appear $\sim$2 mag brighter, although the undetected 
hosts may reduce this difference. They are also at least as bright as, and 
probably by $\sim$1 mag brighter than, the brightest cluster galaxies 
(M(H)$\sim$--26). The luminosities of the FSRQ hosts are intermediate between 
host galaxies of low redshift radio-loud quasars and BL Lac objects 
(M(H)$\sim$--26), and the hosts of high redshift radio-loud quasars 
(M(H)$\sim$--29), in good agreement with current unified models for 
radio-loud AGN, taking into account stellar evolution in the elliptical host 
galaxies. Finally, we find an indicative trend between the host and nuclear 
luminosity for the most luminous FSRQs, supporting the suggestion based on 
studies of lower redshift AGN, that there is a minimum host galaxy luminosity 
which increases linearly with the quasar luminosity.  


\end{abstract}

\section{Introduction}

The determination of the properties of the host galaxies of different AGN 
types is a key tool for our understanding of the AGN phenomenon and for 
unification of different types of AGN. Comparison of AGN properties not 
affected by orientation effects (\eg host galaxies) provides a crucial test 
of the current unified models (Antonucci 1993, Urry \& Padovani 1995). Also, 
it sheds light on the role played by the environment for triggering of 
nuclear activity (Hutchings \& Neff 1992) and on the effect of the AGN on its 
host. 

Flat spectrum radio quasars (FSRQ) form a distinct group from ``normal'' 
steep spectrum radio--loud quasars (RLQ). Most quasars with radio spectral 
index $\alpha_R$ $>$--0.5 (f$_{\nu}\propto\nu^{+\alpha}$) are characterized 
by rapid variability, high and variable polarization and high brightness 
temperatures (Fugmann 1988; Impey \& Tapia 1990; Quirrenbach \etal 1992). 
Moreover, almost all FSRQs in complete samples (Wall \& Peacock 1985) are 
core--dominated radio sources and objects observed at different epochs with 
VLBI display superluminal motion (Padovani \& Urry 1992; Vermeulen \& Cohen 
1994). FSRQs share many observed properties with BL Lac objects and they are 
often grouped together into a common class of blazars. The main difference 
between the two classes is that while FSRQs have strong broad emission lines 
of similar intensity to ``normal'' quasars, emission lines are very weak or 
absent in BL Lacs. How much of the distinction between BL Lacs and FSRQs is 
due to intrinsic properties or a consequence of the classification criteria 
remains unclear, indeed by definition BL Lacs are required to have emission 
line equivalent width smaller than 5 \AA, otherwise they are classified as 
FSRQs.  This selection bias may be responsible for the reported emission line 
differences (Scarpa \& Falomo 1997). On the other hand, based on their 
extended radio emission and evolutionary properties (Stickel \etal 1991; 
Padovani 1992) the two groups of blazars appear different.

The rapid variability, high polarization and high luminosity of blazars are 
usually explained in terms of synchrotron radiation strongly relativistically 
beamed close to our line--of--sight (Blandford \& Rees 1978). This is 
supported by the fact that practically all blazars are luminous and rapidly 
variable $\gamma$--ray sources (\eg von Montigny \etal 1995). If the beaming 
hypothesis is correct, it implies the existence of a more numerous parent 
population of objects intrinsically identical to blazars, but with the jet 
directed away from our line of sight.  In the unified schemes (Barthel 1989; 
Urry \& Padovani 1995), that interpret different classes of objects based on 
geometry, the currently favoured model identifies the parent objects of FSRQs 
with high luminosity lobe-dominated (F--R II) radio galaxies (RG), while the 
low luminosity core-dominated (F--R I) RGs represent the parents of BL Lacs 
(Browne 1983; Ulrich 1989; Padovani \& Urry 1992). A comparison of 
statistical properties of FSRQ and BL Lac samples (Padovani 1992) suggests 
that the two classes of blazars represent similar activity phenomena 
occurring in high-- and low--luminosity early type galaxies, respectively. 
However, there exist potential problems in this simple unification, such as 
the discrepant linear radio sizes of RGs and RLQs, dependence on redshift of 
the ratio of RLQs to RGs, the lack of superluminal motion in RGs, and the 
discrepant radio morphologies of some BL Lacs with respect to F--R I RGs (see 
Urry \& Padovani 1995). 

While some effort has gone to study the properties of BL Lacs (\eg Falomo 
1996; Wurtz, Stocke \& Yee 1996, and references therein), no systematic 
investigation of the host properties of FSRQs has so far been undertaken. In 
this paper we present the first deep high spatial resolution (0.27$''$ 
px$^{-1}$, $\sim1''$ FWHM) near-infrared (NIR) imaging study of the host 
galaxies of a complete sample of FSRQs in the H (1.65 $\mu$m) band. While 
most work on AGN host galaxies has traditionally been done in the optical, 
the NIR wavelengths offer many advantages. Optical emission from quasars is 
strongly dominated by the nuclear source. Often the host galaxies are 
interacting systems and appear irregular in the optical because of tidal 
distortion, star formation and dust emission. The luminosity of the massive 
old stellar population, on the other hand, peaks in the NIR, leading to a 
minimum nucleus/host ratio there. With increasing redshift, one also needs to 
apply much lower K--correction than in the optical. 

The FSRQ sample is taken from the 2Jy (Wall \& Peacock 1985) catalogue of 
radio sources, including all flat spectrum sources at z$<$1.0, at declination 
$\delta$ $\leq$20\degmark ~and not classified as BL Lacs. This yields a total 
of 20 sources. General properties of the objects are given in Table 1. For 
information of the radio and X-ray properties of the sample, see Padovani 
(1992) and Sambruna (1997), respectively. We investigate the properties of 
the host galaxies of FSRQs using 1--D luminosity profile decomposition into 
nuclear and galaxy components. We compare the host absolute magnitudes and 
scale lengths with those of BL Lac hosts, ``normal'' RLQ hosts and RGs, and 
study the relationship between host galaxies and nuclear activity.

\begin{table}
\begin{center}
\begin{tabular}{llccc}
\multicolumn{5}{c}{{\bf Table 1.} The sample.}\\
\hline\hline\\
\multicolumn{1}{c}{Name} & Other name & z & V & M(B) \\
\hline\\
PKS $0208-512 $&         & 1.003 & 16.9 & --26.8 \\
PKS $0336-019 $& CTA 26  & 0.852 & 18.4 & --24.9 \\
PKS $0403-132 $& OF--105 & 0.571 & 17.1 & --25.4 \\
PKS $0405-123 $& OF--109 & 0.574 & 14.9 & --27.7 \\
PKS $0420-014 $& OF--035 & 0.915 & 17.0 & --26.5 \\
PKS $0440-003 $& NRAO 190& 0.844 & 19.2 & --24.3 \\
PKS $0454-463 $&         & 0.858 & 17.4 & --26.5 \\
PKS $0605-085 $& OH--10  & 0.872 & 18.5 & --25.4 \\
PKS $0637-752 $&         & 0.654 & 15.7 & --27.0 \\
PKS $0736+017 $& OI 61   & 0.191 & 16.5 & --23.5 \\
PKS $1055+018 $& 4C 01.28& 0.888 & 18.3 & --25.3 \\
PKS $1226+023 $& 3C 273  & 0.158 & 12.8 & --26.9 \\
PKS $1253-055 $& 3C 279  & 0.538 & 17.7 & --24.6 \\
PKS $1504-166 $& OR--107 & 0.876 & 18.5 & --25.5 \\
PKS $1510-089 $& OR--017 & 0.361 & 16.5 & --25.1 \\
PKS $1954-388 $&         & 0.626 & 17.1 & --25.3 \\
PKS $2128-123 $& OX--148 & 0.501 & 16.1 & --26.1 \\
PKS $2145+067 $& 4C 06.69& 0.990 & 16.5 & --27.4 \\
PKS $2243-123 $&OY--172.6& 0.630 & 16.4 & --26.4 \\
PKS $2345-167 $& OZ--176 & 0.576 & 18.4 & --24.1 \\
\hline       &        &          &   &       \\
\end{tabular}
\end{center}
\end{table}

This paper is arranged as follows. In section 2, we describe the observations 
and data reduction. Section 3 gives the modelling of the profiles, while in 
section 4 we present the derived host parameters and discuss the properties of 
the sample with respect to other classes of AGN. Conclusions are given in 
section 5. In the Appendix, we compare our results for individual objects 
with existing results in the literature. Hubble constant H$_0$ = 50 km 
s$^{-1}$ Mpc$^{-1}$ and deceleration parameter q$_0$ = 0 are used throughout 
this paper. 

\section{Observations and Data Reduction}

We have obtained NIR broad--band images at H (1.65 $\mu$m) band of 20 FSRQs. 
The observations were carried out during two observing runs (in August 1995 
and January 1996) at the ESO/MPI 2.2m telescope at the European Southern 
Observatory (ESO), La Silla, Chile. We used the 256$\times$256 px IRAC2 NIR 
camera (Moorwood \etal 1992) and pixel scale 0.27$''$ px$^{-1}$, giving a 
field of view of 69 arcsec$^2$. Details of the observations and NIR 
photometry are given in Table 2. We shifted the target in a 2$\times$2 grid 
across the array between the observations with typical offsets of 30$''$, 
thus keeping the target always in the field and using the other exposures as 
sky frames. Individual exposures were of 60 sec duration; these were coadded 
to achieve the final integration time. 

\begin{table}
\begin{center}
\begin{tabular}{llrll}
\multicolumn{5}{c}{{\bf Table 2.} Journal of observations.}\\
\hline\hline\\
\multicolumn{1}{c}{Name} & \multicolumn{1}{c}{Date}
    & T$_{int}$ & FWHM   & 6$''$ ap.\\ 
	       &         & min  & arcsec & mag \\
\hline\\
PKS $0208-512 $& 19/8/95 & 6  & 1.2  & 11.76\\ 
    $         $& 21/8/95 & 20 & 1.1  & 11.70\\ 
PKS $0336-019 $& 12/1/96 & 72 & 1.0  & 16.39\\ 
PKS $0403-132 $& 11/1/96 & 60 & 1.2  & 14.97\\ 
PKS $0405-123 $& 13/1/96 & 27 & 0.8  & 13.33\\ 
PKS $0420-014 $& 12/1/96 & 69 & 0.9  & 13.62\\ 
PKS $0440-003 $& 13/1/96 & 75 & 1.0  & 16.04\\ 
PKS $0454-463 $& 11/1/96 & 66 & 1.0  & 15.91\\ 
PKS $0605-085 $& 12/1/96 & 42 & 0.9  & 12.71\\ 
PKS $0637-752 $& 12/1/96 & 72 & 1.0  & 14.79\\ 
PKS $0736+017 $& 11/1/96 & 45 & 0.9  & 13.66\\ 
PKS $1055+018 $& 11/1/96 & 51 & 0.9  & 15.19\\ 
PKS $1226+023 $& 12/1/96 & 20 & 0.9  & 10.92\\ 
PKS $1253-055 $& 21/8/95 & 40 & 2.0  & 12.92\\ 
PKS $1504-166 $& 18/8/95 & 40 & 1.3  & 17.40\\ 
PKS $1510-089 $& 20/8/95 & 40 & 2.0  & 13.29\\ 
PKS $1954-388 $& 19/8/95 & 60 & 0.9  & 14.15\\ 
PKS $2128-123 $& 18/8/95 & 80 & 0.9  & 14.32\\ 
PKS $2145+067 $& 21/8/95 & 60 & 1.1  & 14.47\\ 
PKS $2243-123 $& 18/8/95 & 40 & 0.9  & 14.91\\ 
PKS $2345-167 $& 18/8/95 & 60 & 0.8  & 13.43\\
\hline \\ 
\end{tabular}
\end{center}
\end{table}

Data reduction was performed using IRAF. First, from a raw flat--field frame, 
we marked all bad pixels that were subsequently corrected for in all 
flat--field and science frames by interpolating across neighboring pixels. 
The corrected ON and OFF flat--field frames were subtracted from each other, 
and in case of several ON--OFF pairs, averaged together. The resulting 
flat--fields for each filter and each night were finally normalized to create 
the final flat--fields. For each science frame, a sky frame was produced by 
median averaging all the other frames in a grid. This median sky frame was 
then scaled to match the median intensity level of the science frame, and 
subtracted. Finally, flat--field correction was applied to each 
sky--subtracted frame to produce the final reduced images. All the images of 
the same target were then aligned, using field stars or the centroid of the 
light distribution of the object as a reference point, and combined in order 
to produce the final reduced images that will be used in the subsequent 
analysis. 

Standard stars from Landolt (1992) were observed frequently throughout the 
nights to provide the photometric calibration zero points. We estimate 
photometric accuracy of $\sim$$\pm$0.1 mag. K--correction was applied to the 
host galaxy magnitudes following the method of Neugebauer \etal (1985; their 
Table 3) for a first--ranked elliptical galaxy. The applied K--correction for 
each source is reported in Table 3, column 3. The size of the correction, 
insignificant at low redshift, is m(H)$\sim$0.14 at our median redshift of z = 
0.65. No K--correction was applied to the nuclear quasar component, since for 
a power law spectrum the K--correction equals to (1 + z)$^{1+\alpha}$, where 
$\alpha$$\sim$ --1 for quasars. 

\section{Modelling of the luminosity profiles}

Because of the relatively high redshift of the quasars, the extended emission 
around them is faint and consequently rather noisy. Therefore, in our 
analysis we have considered only the azimuthally averaged fluxes. After 
masking all the regions around the target contaminated by companions, we have 
derived for each object the radial luminosity profile out to a radius where 
the signal was not distinguished from the background noise. This corresponds 
typically to surface brightness of $\mu$(H) = 23--24 mag arcsec$^{-2}$, 
depending on exposure time and observing conditions. Similar procedure was 
followed for the field stars (when available), to obtain the point spread 
function (PSF) suitable for each image. Since the field of view is small, 
only for few objects a star of brightness comparable (or brighter) to the 
target was present in the observed field. For most sources only fainter stars 
were available, therefore extrapolation of the PSF was required at the lowest 
flux levels. This extrapolation is particularly important for marginally 
resolved objects for which the reliability of derived host properties 
strongly depends on the assumed PSF shape. 

We have adopted a functional form to describe the shape of the PSF as a 
Moffat (1969) function, characterized by $\sigma$ for the core and $\beta$ 
for the wings of the PSF. A fit of a Moffat function to the shape of the 
profiles of the available bright stars was found to be quite a good 
representation of the observed PSF. In order to describe the shape of the PSF 
for fields with no stars or only faint stars, we have determined $\sigma$ 
from fitting the core of the stellar profile, whereas $\beta$ was derived 
from the values obtained by fitting bright stars in other frames observed 
during the same night with similar seeing conditions.  For a few objects, no 
stars were visible in the observed field and in these cases the target 
itself, which is always dominated by the nuclear source, was used to estimate 
$\sigma$ and $\beta$ as described above.

The luminosity profiles were fitted into a point source (described by the PSF) 
and a galaxy (described by de Vaucouleurs law, convolved with the PSF) 
components by an iterative least-squares fit to the observed profile. Noisy 
outer parts of the profiles were rejected from the fit. There are three free 
parameters in our fitting: the PSF and bulge intensities at the center, and 
the effective radius of the bulge. We also tried to fit a number of profiles 
with a combination of a PSF and a disk, and although in many cases no 
significant difference was found with respect to an elliptical fit, in no 
case did the disk model yield a better fit. This is not surprising, since 
RLQs are expected to be hosted in early--type galaxies, as well demonstrated 
for low redshift objects by \eg Taylor \etal (1996; hereafter T96) and 
Bahcall \etal (1997). For sources with no host galaxy detected in our 
observations, we determined an upper limit to the brightness of the host 
galaxy by adding simulated ``host galaxies'' of various brightness to the 
observed profile until the simulated host became detectable within the errors 
of the luminosity profile. 

A main problem with the fitting is related to the uncertainty in the sky 
background level. We have checked this by adding or subtracting counts, 
corresponding to 1 $\sigma$ level of the background around the target, from 
the observed profiles, and redoing the fits. The derived host parameters do 
not change much. Another problem is the existence of multiple minima in the 
$\chi^2$--fit, \eg several r(e)--$\mu$(e) pairs can fit the data almost 
equally well. Note that r(e) and $\mu$(e) are expected to be correlated, for 
the total galactic luminosity to be accurately reproduced (see \eg T96). We 
have checked the severity of this problem by starting the fit from various 
different initial values. In general, the fitting program always finds 
roughly the same best values, more easily so for sources with a clearly 
resolved host galaxy. 

We estimate an error in the derived host galaxy magnitudes to be 
$\sim$$\pm$0.3 mag for the clearly detected hosts, this error being largest 
for the sources with the largest nucleus--to--galaxy luminosity ratio. For 
the marginally detected hosts, due to the uncertainties mentioned above, we 
can only assess a lower limit to the error margin as $\geq$$\pm$0.5. 

The fact that we find a very good agreement with previous studies on the 
derived host luminosity for two low redshift FSRQs in our sample (PKS 0736+01 
and PKS 1226+023 = 3C 273; see Appendix) gives us confidence in our adopted 
procedure also for the more problematic derivation of host properties in our 
higher redshift sources. We also note that one of the marginally resolved 
objects, PKS 2128--123 at z = 0.501, has been clearly resolved in the I--band
by the HST (Disney \etal 1995). The host magnitude we derive in the H--band 
is $\sim$1 mag fainter than expected from the I--band for normal galaxy 
colours. However, this difference is not unreasonable, taking into account 
all the uncertainties mentioned above. 

\section{Results and Discussion}

In Fig. 1 we show the H band contour plots of all the FSRQs, after smoothing 
the images with a Gaussian filter of $\sigma$ = 1 px. We detect the host 
galaxy clearly for six (30 \%) FSRQs and marginally for six (30 \%) more. The 
host remains unresolved for eight (40 \%) FSRQs out of 20. We summarize our 
results in Table 3, which gives the best--fit model parameters of the profile 
fitting and the derived properties of the host galaxies. In Fig. 2, we show 
the profiles of each FSRQ, with the best--fit models overlaid. Table 4 
presents a comparison of the H band absolute magnitudes of the FSRQ hosts 
with relevant samples from previous studies in the literature, for which we 
report the average values after correcting the published values for color 
term and to our cosmology (H$_0$ = 50). In the Appendix, we compare our NIR 
photometry with previous existing studies, and discuss in more detail 
individual quasars, including comparison with previous optical/NIR 
determinations of the host galaxies. 

\begin{table*}
\begin{center}
\begin{tabular}{llllrrrlrl}
\multicolumn{10}{c}{{\bf Table 3.} Properties of the host galaxies.}\\
\hline\hline\\
Name & z & K--corr. & r(e)/R(e) & \multicolumn{2}{c}{m$_H$} & L(n)/L(g) & 
\multicolumn{2}{c}{M$_H$} & Note$^*$\\
 & & & ($''$/kpc) & nucleus& host & & nucleus & host & \\
\hline \\
PKS $0208-512$ & 1.003 & 0.30 & 1.85/20.2 & 11.8 & 15.0    & 19.0   & --33.2 
& --30.3    & M\\
PKS $0336-019$ & 0.852 & 0.22 &           & 16.2 & $>$16.6 & $>$1.4 & --28.2 
& $>$--28.0 & U\\
PKS $0403-132$ & 0.571 & 0.12 & 2.75/23.8 & 14.9 & 18.8    & 32.3   & --28.8 
& --25.1    & M\\
PKS $0405-123$ & 0.574 & 0.12 & 1.20/10.4 & 13.4 & 15.6    & 8.1    & --30.0 
& --27.9    & R\\
PKS $0420-014$ & 0.915 & 0.25 & 2.45/25.9 & 13.5 & 17.0    & 24.0   & --31.4 
& --28.1    & R\\
PKS $0440-003$ & 0.844 & 0.21 &           & 16.0 & $>$17.5 & $>$4.0 & --28.3 
& $>$--27.0 & U\\
PKS $0454-463$ & 0.858 & 0.22 &           & 15.9 & $>$18.1 & $>$7.6 & --28.5 
& $>$--26.5 & U\\
PKS $0605-085$ & 0.872 & 0.23 &           &      &         &        &       
&          & U\\
PKS $0637-752$ & 0.654 & 0.14 & 1.35/12.5 & 14.6 & 17.8    & 19.0   & --29.2 
& --26.1    & M\\ 
PKS $0736+017$ & 0.191 & 0.01 & 0.75/~3.2 & 14.3 & 14.3    & 1.0    & --26.2 
& --26.2    & R\\
PKS $1055+018$ & 0.888 & 0.23 &           & 15.2 & $>$17.7 & $>$10.0& --29.3 
& $>$--27.0 & U\\
PKS $1226+023$ & 0.158 & 0.01 & 4.50/16.6 & 10.9 & 13.3    & 9.0    & --29.4 
& --27.0    & R\\
PKS $1253-055$ & 0.538 & 0.12 &           & 12.9 & $>$13.7 & $>$2.1 & --29.6 
& $>$--29.0 & U\\
PKS $1504-166$ & 0.876 & 0.23 &           & 17.6 &         & $>100$ & --26.9 
&          & U \\
PKS $1510-089$ & 0.361 & 0.04 &           & 14.8 & $>$16.2 & $>$3.6 & --27.3 
& $>$--26.0 & U \\
PKS $1954-388$ & 0.626 & 0.14 & 0.65/~5.9 & 14.1 & 16.3    & 7.3    & --29.4 
& --27.3    & R\\
PKS $2128-123$ & 0.501 & 0.11 & 2.00/16.0 & 13.9 & 18.0    & 44.0   & --29.0 
& --25.1    & M\\
PKS $2145+067$ & 0.990 & 0.29 & 1.05/11.4 & 14.4 & 17.7    & 19.0   & --30.5 
& --27.5    & M\\
PKS $2243-123$ & 0.630 & 0.14 & 0.80/~7.3 & 14.8 & 18.2    & 24.0   & --28.9 
& --25.5    & M\\ 
PKS $2345-167$ & 0.576 & 0.13 & 0.90/~7.8 & 13.3 & 15.7    & 9.0    & --30.1 
& --27.8    & R\\
\hline \\
\multicolumn{10}{l}{$^*$: R = resolved, M = marginally resolved, U = unresolved.} 
\end{tabular}
\end{center}
\end{table*}

\begin{table*}
\begin{center}
\begin{tabular}{lcrllll}
\multicolumn{7}{c}{{\bf Table 4.} Comparison of the average host galaxy 
properties with other samples.}\\
\hline\hline \\
\multicolumn{1}{c}{Sample} & 
filter & N & $<z>$ & $<M_B>$  & $<M_H(nuc)>$ & $<M_H(host)>^b$ \\
\hline \\
L$^*$ Mobasher \etal (1993) & K & 136 & 0.077$\pm$0.030 &    & & 
--25.0$\pm$0.2 \\
                            &   &     &     &     & &                \\
BCM Thuan \& Puschell (1989) & H & 84 & 0.074$\pm$0.026 & & & --26.3$\pm$0.3 \\
                            &   &     &     &     & &                \\
RLQ McLeod \& Rieke (1994a) & H & 22 & 0.103$\pm$0.029 &  & --25.1$\pm$0.5 & 
--24.9$\pm$0.6 \\
RLQ McLeod \& Rieke (1994b) & H & 23 & 0.196$\pm$0.047 &  & --26.5$\pm$0.9 & 
--25.7$\pm$0.6 \\
RLQ Bahcall \etal (1997) & V & 6 & 0.220$\pm$0.047 & --25.5$\pm$0.9  & & 
--26.1$\pm$0.5 \\
RLQ Taylor \etal (1996) & K & 13 & 0.236$\pm$0.046 & --24.5$\pm$0.8 & 
--27.1$\pm$0.8 & --26.3$\pm$0.7 \\
RLQ Veron-Cetty \& Woltjer (1990) & I & 20 & 0.343$\pm$0.094 & --25.2$\pm$0.5 
& & --26.3$\pm$0.5 \\
RLQ Hooper \etal (1997) & R & 6 & 0.465$\pm$0.032 &  & --26.8$\pm$0.4 & 
--26.2$\pm$0.4 \\
RLQ R\"{o}nnback \etal (1996) & R & 9 & 0.594$\pm$0.120 & --24.7$\pm$1.1  & & 
--25.8$\pm$0.4 \\
RLQ Lehnert \etal (1992) & K & 6 & 2.342$\pm$0.319 &  & 
--30.5$\pm$1.0 & --28.8$\pm$1.1 \\
                            &   &     &     &     & &                \\
RG Taylor \etal (1996) & K & 12 & 0.214$\pm$0.049 & --21.7$\pm$0.6  & 
--25.1$\pm$0.7 & --26.1$\pm$0.8 \\
                            &   &     &     &     & &                \\
BL Falomo (1996), Wurtz \etal (1996) & R & 48 & 0.194$\pm$0.101 & & 
--25.2$\pm$2.4 & --26.3$\pm$0.7 \\
BL Falomo \etal (1997) & I & 7 & 0.422$\pm$0.186 & & --27.0$\pm$0.8 & 
--26.7$\pm$0.8\\
                    &   &     &     &     & &                \\
FSRQ/R+M$^a$ (0.5$<$z$<$1.0) & H & 9 & 0.671$\pm$0.157 & --26.2$\pm$1.1 & 
--29.7$\pm$0.8 & --26.7$\pm$1.2 \\
FSRQ/R$^a$ (0.5$<$z$<$1.0) & H & 4 & 0.673$\pm$0.141 & --25.9$\pm$1.3  & 
--30.2$\pm$0.7 & --27.8$\pm$0.3 \\
\hline\\
\multicolumn{7}{l}{$^a$: R = resolved; M = marginally resolved.}\\
\multicolumn{7}{l}{$^b$: Transformation of magnitudes to H band done assuming 
V--H = 3.0, R--H = 2.5 and H--K = 0.2 galaxy colours. }\\
\multicolumn{7}{l}{~~~All magnitudes have been converted to our adopted 
cosmology (H$_0$ = 50 \kmpspMpc ~and q$_0$ = 0).} 
\end{tabular}
\end{center}
\end{table*}

\begin{figure*}
\psfig{file=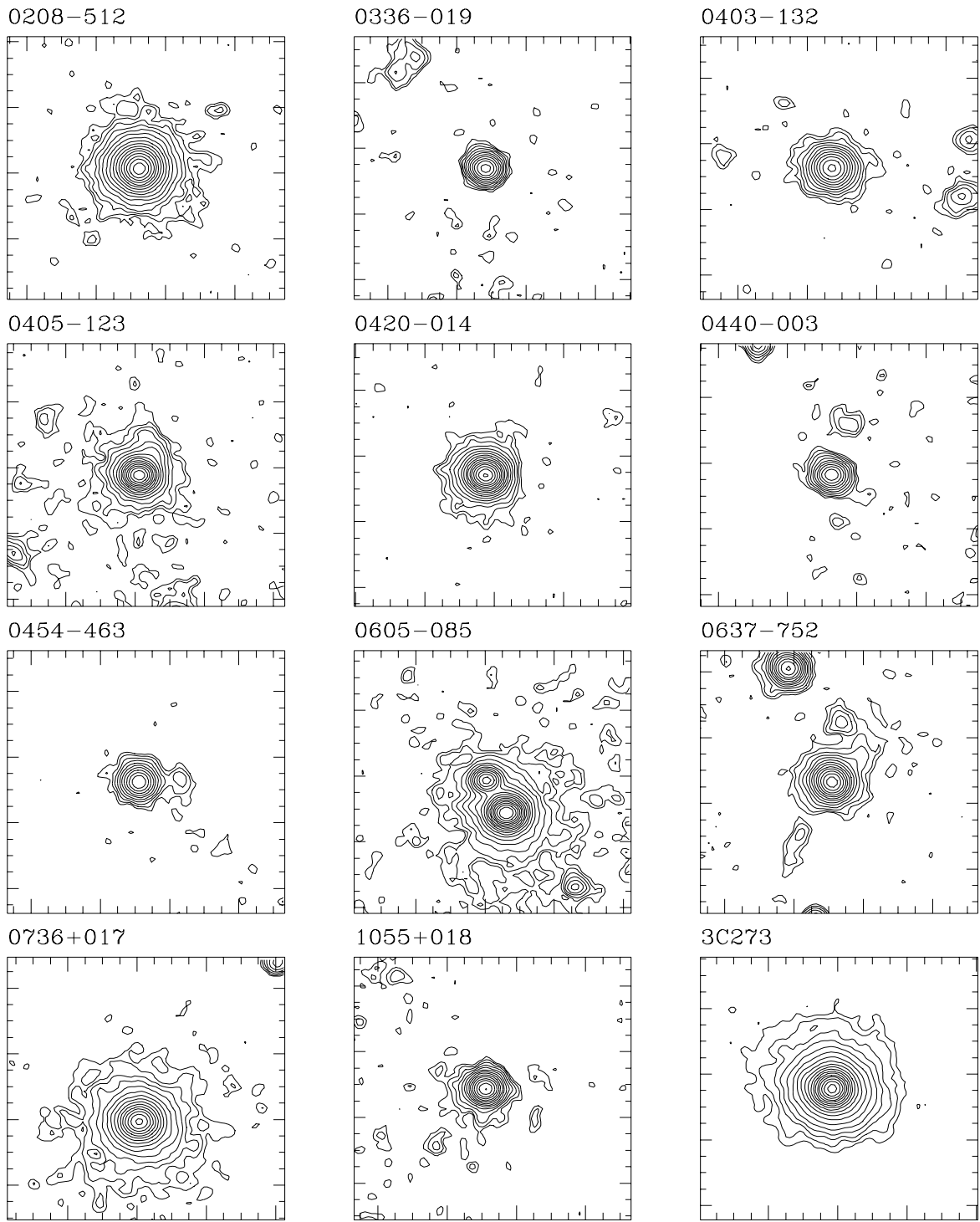,width=18cm,height=22.8cm}
\caption{\label{fig:fig1} 
Gaussian ($\sigma$ = 1 px; 0.27$''$) smoothed contour plots of the sample 
objects in the H band. The full size of the image is 80 px (21.6$''$) across. 
The contours are separated by 0.5 mag intervals. North is up and east to the 
left.
 }
\end{figure*}

\begin{figure*}
\psfig{file=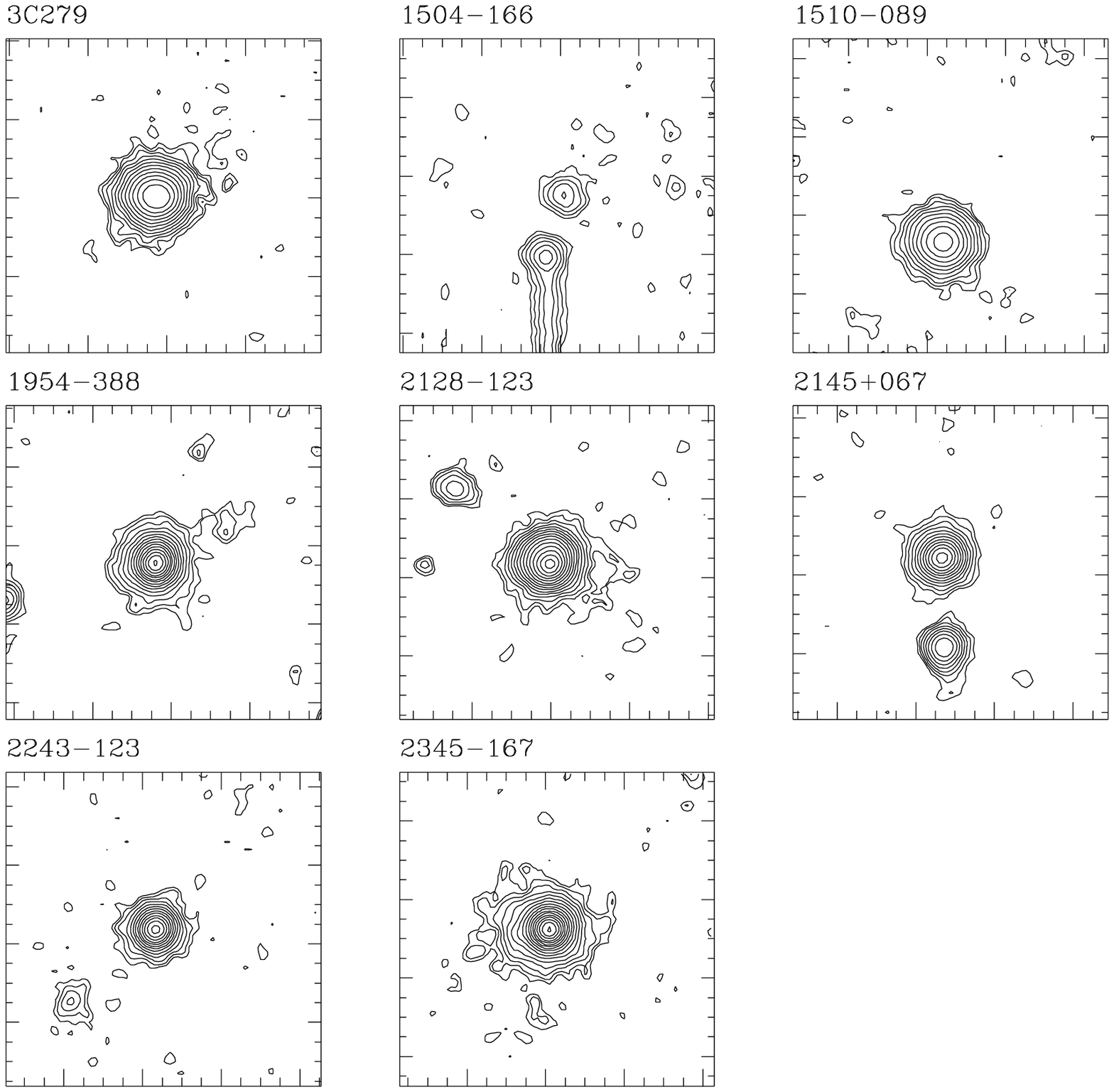,width=18cm,height=22.8cm}
\end{figure*}

\begin{figure*}
\psfig{file=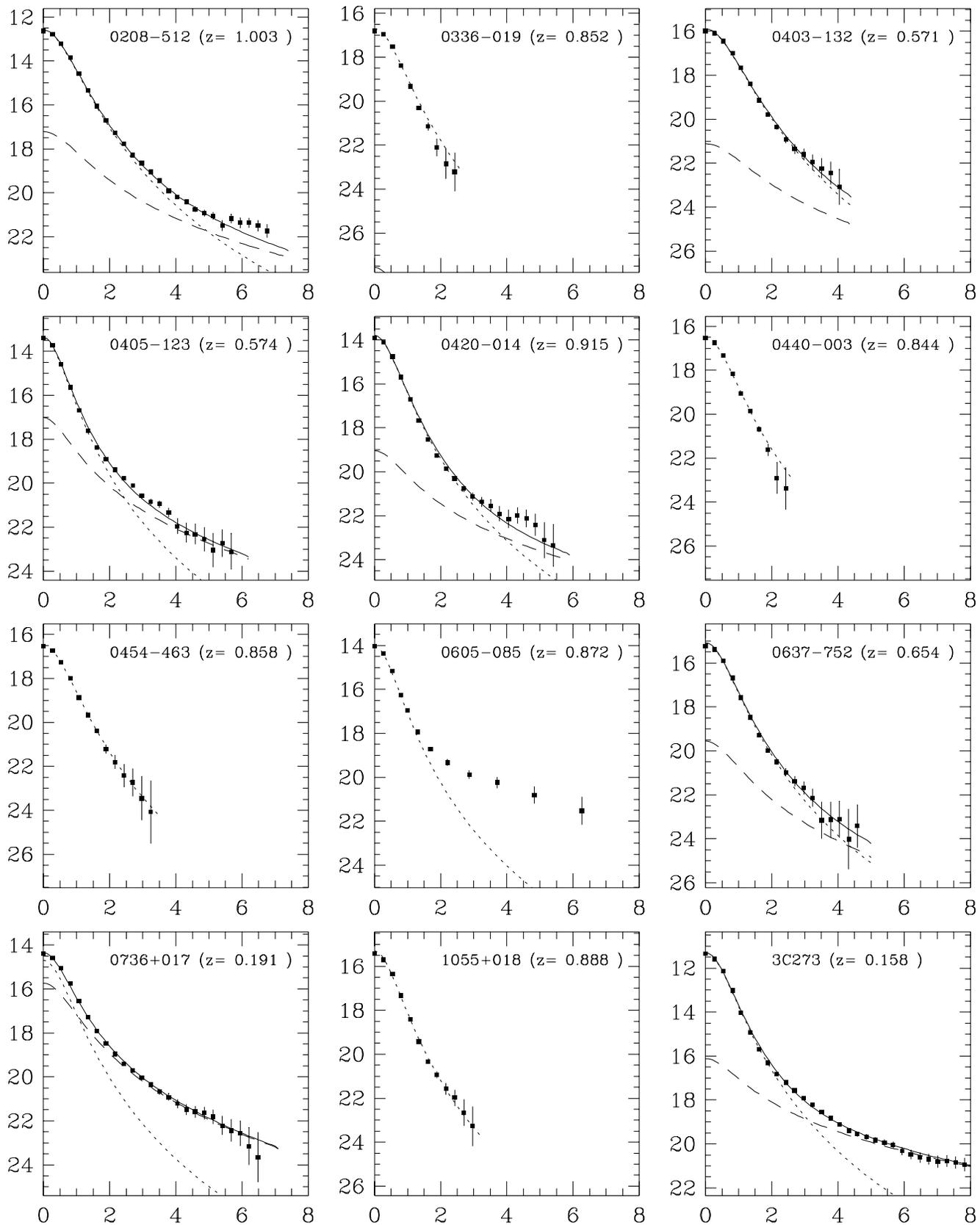,width=18cm,height=22.8cm}
\caption{\label{fig:fig2} 
Results of the profile fits for each galaxy. The solid points represent the 
observed profile, short--dash line the PSF, long--dash line the bulge, and 
the solid line the total theoretical profile. 
}
\end{figure*}

\begin{figure*}
\psfig{file=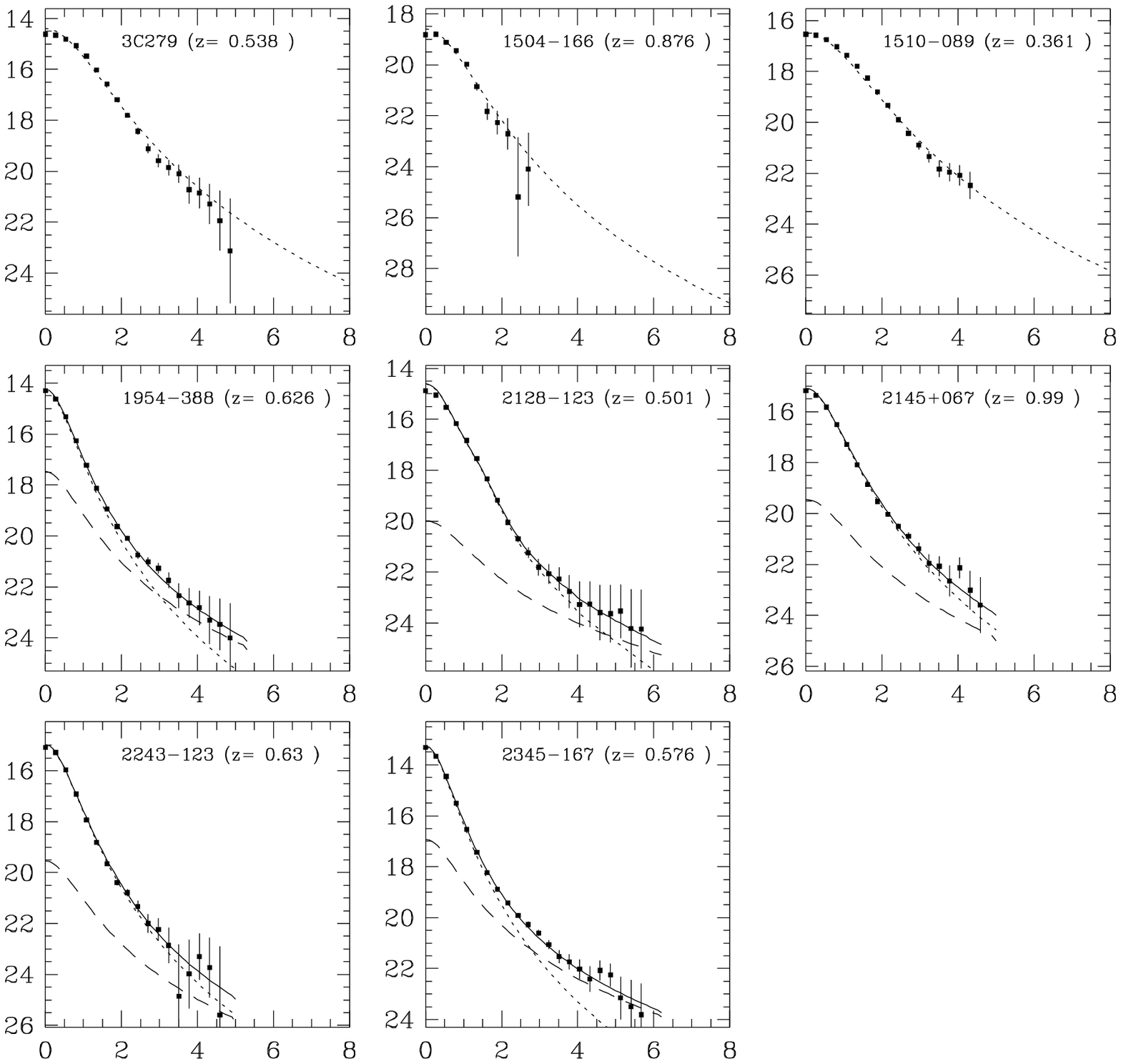,width=18cm,height=17.3cm}
\end{figure*}

\subsection{Host luminosity}

In Fig. 3 (upper panel) we investigate the location of the FSRQ hosts and the 
hosts of various other AGN samples imaged in the NIR in the apparent 
magnitude vs. redshift H--z Hubble diagram, relative to the established 
relation for RGs (solid line, \eg Lilly \& Longair 1984; Lilly, Longair \& 
Allington--Smith 1985; Eales \etal 1997). For comparison, we also show the 
evolutionary model for elliptical galaxies derived from passive models of 
stellar evolution by Bressan, Chiosi \& Fagotto (1994; dashed line, 
normalized to the average redshift and magnitude of the T96 low redshift 
RGs). The resolved FSRQ hosts lie remarkably well on the H--z relation, 
whereas there is large scatter for the marginally resolved FSRQ hosts. In 
Fig. 3 (lower panel) we show the H--z diagram for the mean value of various 
samples of AGN taken from the literature. T96 found for nearby RLQ and RG 
hosts (after removing the nuclear component) that they lie above the 
established RG relation, \ie toward fainter magnitude by $\sim0.5$ mag on the 
average. It is apparent that the same holds true for most other low redshift 
AGN samples and, at intermediate redshift, for the marginally resolved FSRQ 
hosts and for the RLQ hosts of Hooper, Impey \& Foltz (1997). This may 
indicate a continuation, and even strengthening, of the deviation found by 
T96 to higher redshift which is likely explained by the (uncorrected) 
contribution of the nuclear source in RGs (see T96). 

\begin{figure}
\psfig{file=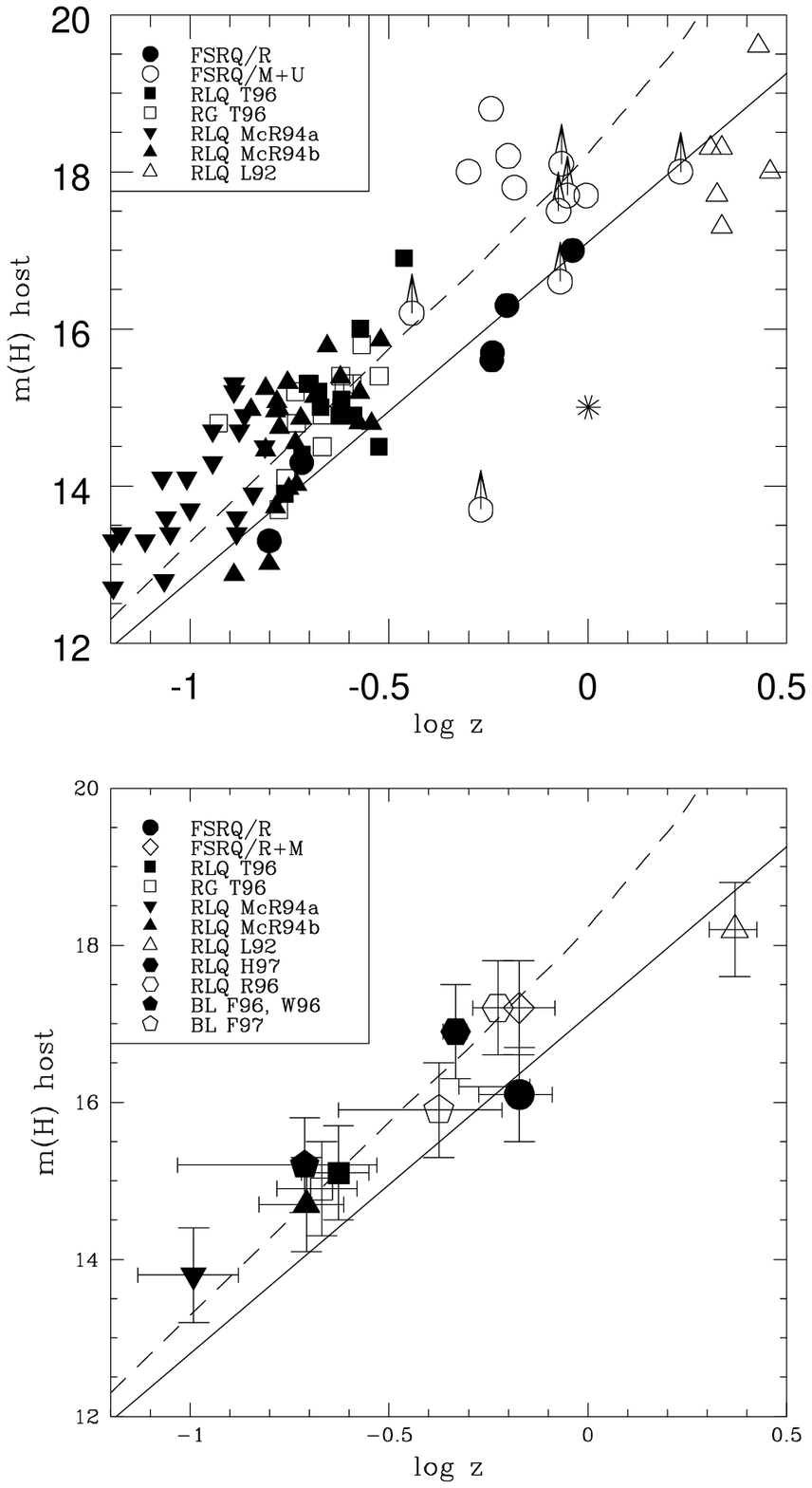,width=8cm,height=16cm}
\caption{\label{fig:fig3} 
{\bf Upper panel:} The apparent magnitude of the host galaxies vs. redshift 
(Hubble diagram). Resolved FSRQs are marked as filled circles, marginally 
resolved FSRQs as open circles and derived limits for the hosts of unresolved 
FSRQs as open circles with arrow. PKS 0208--512 is marked as an asterisk (see 
Appendix). Sources from T96 are marked as filled (RLQ) and open (RG) squares, 
RLQs from McLeod \& Rieke (1994a,b) as solid and inverted solid triangles, and 
z$\sim$2 RLQs from Lehnert \etal (1992) as open triangles. The solid line is 
the Hubble relation for RGs (Lilly \etal 1985; Eales \etal 1997). The dashed 
line is the evolutionary model for elliptical galaxies (Bressan \etal 1994), 
normalized to the average redshift and magnitude of the low redshift RGs of 
T96.
{\bf Lower panel:} As in the upper panel, except for the mean values of the 
FSRQs in comparison with samples from literature. The diamond represents the 
combined sample of resolved and marginally resolved FSRQs, excluding the two 
low redshift objects (PKS 0736+017 and 3C 273). Additional samples from 
optical imaging by R\"{o}nnback \etal (1996, RLQ), Hooper \etal (1997, RLQ), 
Falomo (1996, BL Lacs), Wurtz \etal (1996, BL Lacs) and Falomo \etal (1997, 
BL Lacs) are indicated as marked in the figure. 
}
\end{figure}

Interestingly, however, the firmly detected FSRQ hosts (and the intermediate 
redshift RLQ hosts of R\"{o}nnback \etal 1996) have relatively brighter 
magnitudes than the low redshift RLQs and the intermediate redshift RLQs of 
Hooper \etal (1997), consistent with the established RG relation (solid line 
in Fig. 3). Also, the magnitudes of the high redshift (z$\sim$2) RLQ hosts 
studied by Lehnert \etal (1992) are consistent with the H--z relation, 
suggesting that between redshifts of z$\sim$0.5 and z$\sim$2 there is an 
increase in the host brightness with respect to the Hubble diagram. Similar 
result for high redshift RGs was noted by Eales \etal (1997) who found that 
while the Hubble diagram of luminous 3C RGs and fainter 6C/B2 RGs were 
similar at z$<$0.6, the 3C RGs are $\sim$0.6 mag brighter at z$>$1. Eales 
\etal attribute this change not to stellar evolution, but to a difference in 
the intrinsic luminosity of the AGN component of the RG samples studied. 
There is also an obvious selection effect in that at very high redshifts, 
only very bright host galaxies can be detected (see also Fig. 4). 

In Fig. 4, we show the H band absolute magnitude vs. redshift for the host 
galaxies from various samples. The average H band absolute magnitude of the 
resolved FSRQ host galaxies is M(H) = --27.4$\pm$0.6 and the average bulge 
scale length 11.6$\pm$7.6 kpc, while the values after adding the marginally 
resolved hosts are --26.7$\pm$1.2 and 12.8$\pm$6.0 kpc. The absolute 
magnitude considering only the four resolved FSRQs at 0.5$<$z$<$1.0 is M(H) = 
--27.8$\pm$0.3. The FSRQ hosts are therefore large (all have R(e)$>$3 kpc, 
the empirical upper boundary found for normal local ellipticals by Capaccioli, 
Caon \& D'Onofrio 1992), and very luminous, much brighter than the 
luminosity of an L$^*$ galaxy, which has M(H) = --25.0$\pm$0.3 (Mobasher, 
Sharples \& Ellis 1993). It is therefore evident that the clearly detected 
FSRQ hosts are preferentially selected from the high--luminosity tail of the 
galaxy luminosity function (the derived upper limits for the unresolved hosts 
are also consistent with this). Indeed, we find no case of an FSRQ host with 
M(H)$>$--25, indicating that for some reason these quasars cannot be hosted 
by a galaxy with L$<$L$^*$ (similarly to what was found by T96). The FSRQ 
hosts have also slightly brighter luminosities than the mean value of 
brightest cluster member galaxies (BCM; M(H) = --26.3$\pm$0.3; Thuan \& 
Puschell 1989), although there are several FSRQ hosts that fall into the BCM 
range.

\begin{figure}
\psfig{file=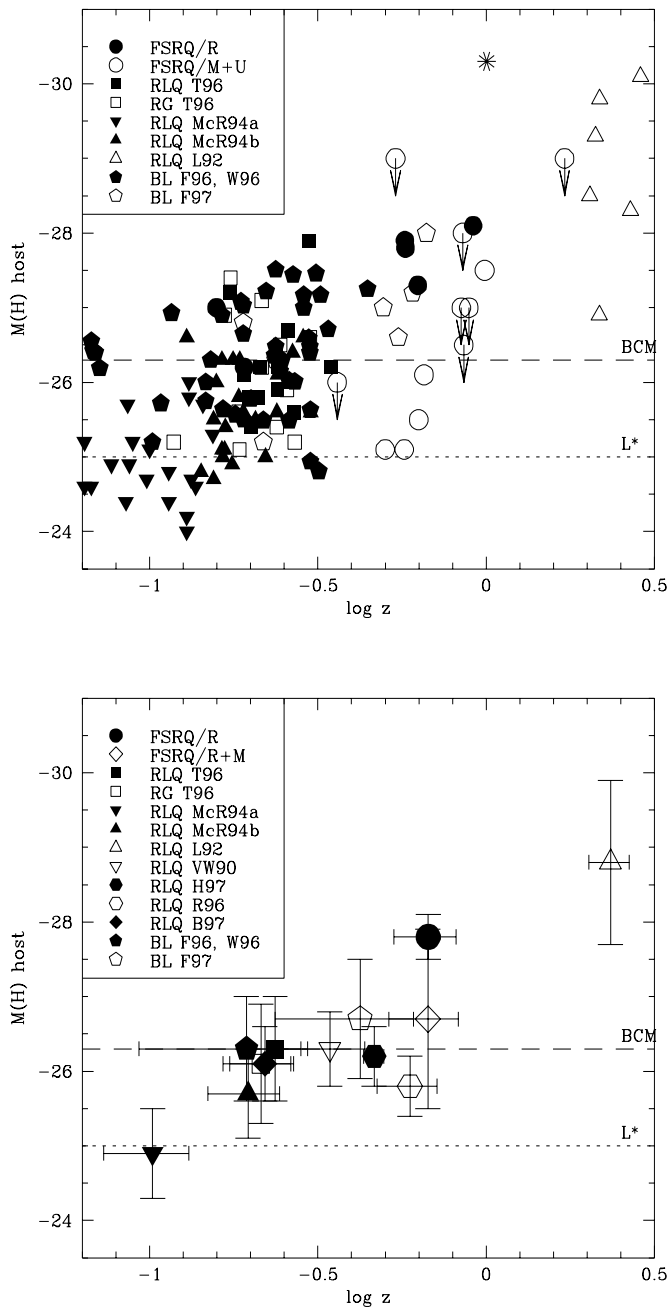,width=9cm,height=18cm}
\caption{\label{fig:fig4}
{\bf Upper panel:} Plot of the absolute H band magnitude of the host galaxies 
vs. redshift. The average luminosities of L* galaxies (M(H)$\sim$--25.0; 
Mobasher \etal 1993) and brightest cluster member galaxies (BCM; 
M(H)$\sim$--26.3; Thuan \& Puschell 1989) are indicated as long--dash and 
short--dash lines,respectively. For symbols, see Fig. 3.
{\bf Lower panel:} As the upper panel, except for the mean values of various 
samples. Additional samples based on optical imaging from Veron--Cetty \& 
Woltjer (1990, RLQ) and Bahcall \etal (1997, RLQ) are indicated as marked in 
the figure. For other symbols, see Fig. 3. 
}
\end{figure}

Most of the available comparison data are for low and intermediate redshift 
RLQs. The samples we have retrieved from literature span a moderately large 
range in redshift from z$\sim$0.1 up to z$\sim$0.6, slightly smaller than the 
average redshift of the FSRQ sample. The average host galaxy magnitudes for 
the various samples are given in Table 4. The RLQ samples we consider are (in 
order of increasing average redshift) from McLeod \& Rieke (1994a,b), Bahcall 
\etal (1997), T96, Veron--Cetty \& Woltjer (1990), Hooper \etal (1997) and 
R\"{o}nnback \etal (1996). Considering all these samples together gives 
average host magnitude of M(H) = --25.9$\pm$0.4. As can be seen from Fig. 4, 
there is no significant difference between the average values of these samples.
Considering first conservatively both the resolved and marginally resolved 
FSRQ hosts gives average host magnitude of M(H) = --26.7$\pm$1.2, \ie slightly 
brighter but $<$1$\sigma$ away from the average RLQ value. On the other hand, 
considering only the firmly detected FSRQ hosts with z$>$0.2, the average 
M(H) = --27.8$\pm$0.3, more significantly brighter but still consistent with 
the low--z RLQs within 3$\sigma$. The simplest unified model states that all 
RLQs are similar; it is therefore not surprising that RLQ and FSRQ hosts are 
reasonably similar, especially considering the small number of sources 
analyzed. However, the persistent 1--2 magnitude difference of FSRQ hosts 
with AGN hosts at lower redshift suggests evolution in the host brightness 
with redshift, and/or a relationship of the host luminosity with the nuclear 
luminosity (see section 4.2). 

Lehnert \etal (1992) reported spatially resolved structures in the K band 
around six RLQ at z$\sim$2.3 that, if interpreted as host galaxies, would 
correspond to extremely luminous galaxies (average host M(H) = 
--28.8$\pm$1.1), $\sim$1--2 mag brighter than the FSRQs at z$\sim$0.65. 
However, within the scatter involved in these numbers, our results appear to 
be consistent with those of Lehnert \etal (1992), both for the evolutionary 
trend in the Hubble diagram (see above) and for the trend between the nuclear 
and host galaxy luminosities (see section 4.2.), and is supporting evidence 
for the existence of a real upturn in the host luminosity occurring between 
z$\sim$0.5 and z$\sim$2, leading from L$\geq$L* hosts at low redshift to the 
host galaxies of high redshift quasars that are several magnitudes brighter 
than L* (see Fig. 4). While this type of change is consistent with evolution 
of the stellar population in the elliptical hosts (as argued for high 
redshift RGs by Lilly \& Longair 1984), or being intrinsic AGN luminosity 
effect (as argued for high redshift RGs by Eales \etal 1997), there are many 
caveats in this comparison, most notably differences in the intrinsic quasar 
luminosity of the various samples. In addition, optical and NIR imaging by 
Lowenthal \etal (1995) failed to detect extended emission in a sample of six 
radio--quiet quasars (RQQ) at z$\sim$2.3. Their upper limits indicate that 
the RQQ hosts at high redshift must be $\leq$3 mag brighter than L* and 
$\geq$1 mag fainter than the Lehnert (1992) sample of RLQs at similar 
redshift, suggesting that RLQs and RQQs are different types of objects. 

While other explanations for light around high redshift RLQs have been 
proposed, \eg foreground galaxies producing intervening MgII 2800 
\AA ~absorption lines (LeFevre \& Hammer 1988) or light from a hidden quasar 
scattered by dust or electrons along the radio axis (Fabian 1989), starlight 
from a host galaxy remains the most likely alternative, given that high 
redshift RGs can reach similar luminosities and the quasar nebulosities 
follow remarkably well the tight Hubble diagram for RGs (\eg Lilly 1989; 
Eales \etal 1997). 

There has been considerable disagreement on the similarity between the hosts 
of RGs and RLQs. While some authors have found similar size and morphology 
(\eg Barthel 1989, Veron--Cetty \& Woltjer 1990, Lehnert \etal 1992), others 
have concluded that RLQ hosts are brighter by 0.5--1.0 mag than RGs of 
similar extended radio emission (\eg Smith \etal 1986; Hutchings 1987; Smith 
\& Heckman 1989). Abraham, Crawford \& McHardy (1992) showed that this 
disagreement is most likely due to underestimation of RLQ host luminosity due 
to difficulties in PSF subtraction (because of cosmological host surface 
brightness dimming and scattered light from the nuclear component), and that 
RLQ hosts are in fact brighter than RGs. However, using carefully matched 
samples, T96 found that RLQ and RG hosts are almost identical in morphology, 
scale length and luminosity, and moreover, the nuclear components of RGs are 
fainter and redder than those in RLQS, all in good agreement with the unified 
model. At average redshift z = 0.214$\pm$0.049, the average host magnitude of 
the host galaxies of RGs in the study of T96 is M(H) = --26.1$\pm$0.8. This 
value agrees reasonably well with the higher redshift FSRQ hosts, taking into 
account some stellar evolution in the early type host galaxies. The data 
presented in this paper therefore support the similarity between RLQ and RG 
hosts, considering also the good agreement between the magnitudes of the FSRQ 
host galaxies and the high redshift RGs used to produce the Hubble diagram 
(Fig. 3; see Eales \etal 1997). 

FSRQs share many properties (\eg variability and polarization) with BL Lac 
objects and it is therefore interesting to compare the host properties of 
these two classes of blazars. Recent optical R band investigations of BL Lac 
hosts at z$\leq$0.5 by Falomo (1996) and Wurtz \etal (1996) find the average 
absolute magnitude of the host galaxies to be M(H)$\sim$--25.8 (see Table 4), 
with some indication of positive correlation of host brightness with 
increasing redshift. HST R band imaging of a small number of BL Lacs at 
z$>$0.5 (Falomo \etal 1997) has provided additional evidence for more 
luminous hosts (M(H)$\leq$--26.8) at higher redshift. Although based on a 
small number of resolved objects, it appears therefore that, accounting for 
stellar evolution that makes galaxies brighter by $\sim$1 mag between z 
= 0 and z = 1, the hosts of FSRQs have similar luminosity to lower redshift 
BL Lacs, in agreement with the unified model. Note also that recent 
spectroscopic study of the emission line properties (Scarpa \& Falomo 1997), 
which is one of the main distinctive characteristics between FSRQs and BL 
Lacs, yields additional support to this scenario.

\subsection{The nuclear component}

The average absolute magnitude of the fitted nuclear component for all FSRQs 
is M(H) = --29.7$\pm$0.8. This indicates that FSRQ nuclei are on average 
$\sim$2.5--3 mag brighter than the RLQ nuclei at lower redshift (\eg T96; 
M(H) = --27.1$\pm$0.8) and $\sim$4.5--5 mag brighter than the nuclear 
components in low redshift RGs (\eg T96; M(H) = --25.1$\pm$0.7). The presence 
of a strong nuclear component in FSRQ is even more evident when considering 
the nucleus/galaxy (N/G) luminosity ratio, shown in Fig. 5. None of the low 
redshift RLQs and RGs studied by McLeod \& Rieke (1994b) and T96 have N/G 
$>$10 in the H band, whereas about half of our FSRQs are above this limit. 

\begin{figure}
\psfig{file=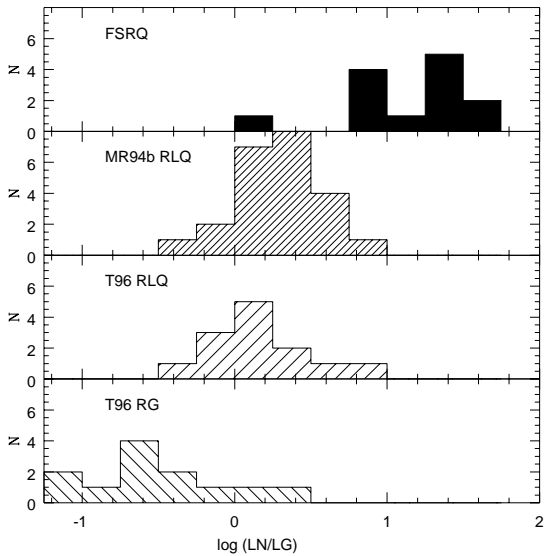,width=9cm,height=9cm}
\caption{\label{fig:fig5}
Histogram of the nucleus/host luminosity ratio for the FSRQs, and for the low 
redshift RLQs from McLeod \& Rieke (1994b) and low redshift RLQs and RGs from 
T96. The K--band data from T96 has been converted into the H--band assuming 
colour for the nuclear and galaxy components of H--K = 1.1 and 0.2, 
respectively. 
}
\end{figure}

From Figs. 3 and 4 it appears  that the host galaxies of the various control 
samples considered  here are not dramatically different in intrinsic 
luminosity, especially if some stellar evolution in the elliptical host 
galaxies is taken into account. Therefore, Fig. 5 clearly indicates that 
FSRQs exhibit a nuclear component which is systematically brighter than that 
of other AGN. This is consistent with the beaming model with large Doppler 
amplification factor that makes the observed difference of $\sim$3 magnitudes 
understandable. 

In Fig. 6, we show the relation between the luminosities of the nucleus and 
the host galaxy for the FSRQs, and for various samples from literature, for 
individual quasars (upper panel) and for the mean values of the samples (lower 
panel). While T96 found no convincing correlation between the host and AGN 
luminosity, we find there is a tendency for the more powerful FSRQs to reside 
in more luminous hosts. Similar trend has previously been noted in the NIR 
for low redshift quasars (McLeod \& Rieke 1994a,b) and for Seyfert galaxies 
(Danese \etal 1992, Kotilainen \& Ward 1994). Moreover, recent optical 
observations of bright (M(R)$<$--24) quasars at 0.4$<$z$<$0.5 (Hooper \etal 
1997) also indicate what the authors call a positive correlation between the 
host and nuclear luminosity. 

\begin{figure}
\psfig{file=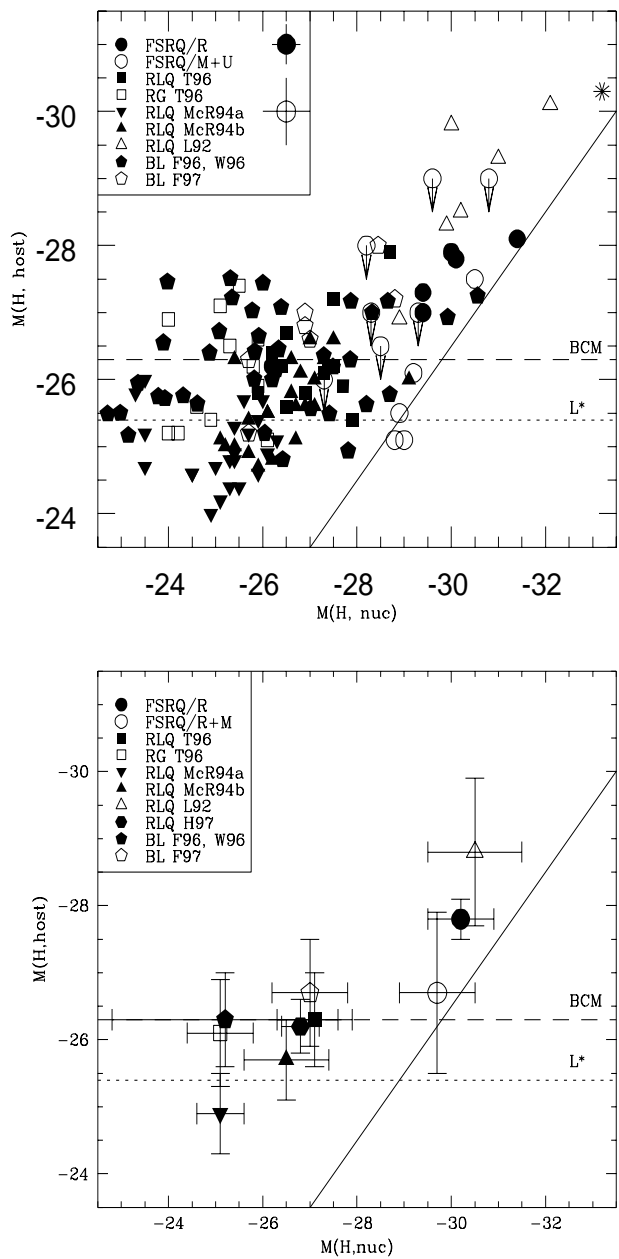,width=9cm,height=18cm}
\caption{\label{fig:fig6} 
{\bf Upper panel:} 
Plot of the H band nuclear vs. host luminosity. For symbols, see Fig. 3 and 
4. The solid line is the limiting mass--luminosity envelope from the M(B,nuc) 
vs. M(H,host) diagram of McLeod \& Rieke (1995), converted to H band using a 
least squares fit of the M(B,nuc) and M(H,nuc) values for the FSRQs from 
Tables 1 and 3. The two large circles represent the estimated error in the 
derived host galaxy magnitudes ($\sim$$\pm$0.3 mag for the clearly detected 
hosts; filled circle, and $\geq$$\pm$0.5 for the marginally detected hosts; 
open circle).
{\bf Lower panel:} As in the upper panel, except for the mean values of 
various samples. For symbols, see Fig. 3 and 4.
}
\end{figure}

Note that not only the fully and marginally resolved FSRQ hosts, but also all 
the upper limits derived for the unresolved hosts are well consistent with 
the boundary limit proposed by McLeod \& Rieke (1995) for AGN with 
M(B)$<$--23 (solid line in Fig. 6). They interpret the limit in the sense 
that there is a minimum host galaxy luminosity which increases linearly with 
quasar luminosity. Recently, McLeod (1997) has speculated that this 
relationship represents a constant ratio of the central black hole mass to the 
host galaxy mass. Finally, we advise caution about possible selection effects 
in this relationship.  Since faint host galaxies are difficult to be detected 
under the most luminous nuclei, we may expect that this contributes to the 
void of sources in the lower right--hand corner in Fig. 6. On the other hand, 
in the case of beamed objects such as FSRQs the effects of amplification of 
the nuclear source may move systematically the points towards larger nuclear 
luminosities.

\section{Summary and Conclusions}

The main finding of our NIR study is that we can resolve the host galaxies of 
a significant fraction of luminous AGN out to considerable redshift. The host 
galaxies of z$\sim$0.65 FSRQs are large (average bulge scale length 
$\sim$13$\pm$7 kpc) and bright (average M(H)$\sim$--27$\pm$1), much more 
luminous than L$^*$ galaxies (by $\sim$2 mag) and somewhat more luminous than 
the brightest cluster galaxies (by $\sim$1 mag). Note that all detected hosts 
have M(H)$<$--25 ($\sim$L$^*$) and the derived upper limits are consistent 
with this value. The FSRQ hosts are 1--2 mag brighter than the hosts of lower 
redshift RLQs, and $\sim$1 mag fainter than the hosts of z$\sim$2 RLQs, 
consistent with stellar evolution in the elliptical host galaxies and unified 
models. Finally, the FSRQ hosts appear $\sim$1 mag brighter than the hosts of 
lower redshift BL Lac objects, again consistent with them forming a common 
class of blazars, if mild stellar evolution in their host galaxies is assumed. 

The luminosity of the host shows a positive trend with that of the active 
nucleus, at least for the most luminous sources. This enforces the suggestion 
that, for the brightest AGN, there is a minimum host galaxy luminosity which 
increases linearly with quasar luminosity.  However, since several objects 
remain unresolved, deeper and higher resolution NIR imaging is required for 
these sources in order to determine their host properties.

\section*{Acknowledgments} JKK acknowledges a research grant from the Academy 
of Finland during the initial part of the course of this work.

\section*{Appendix: Notes on individual objects and comparison with previous 
NIR photometry}

{\bf PKS 0208--512}. The H band image of this z = 1.003 FSRQ shows quite a 
smooth morphology. The host galaxy is marginally resolved. Unfortunately, this 
source has no reference stars in the observed field. The profile fit shown in 
Fig. 2 has been derived using a PSF estimated from field stars in the frames 
taken as close as possible in time and with similar seeing conditions. With 
these assumptions, the implied host galaxy is the most luminous in the FSRQ 
sample (M(H) = --30.3). However, due to the uncertainty in the PSF shape, we 
have omitted this FSRQ from the statistical analysis and discussion. No 
previous NIR photometry was found for PKS 0208--512.

{\bf PKS 0336--019}. This source at z = 0.852 remains unresolved. Our H band 
magnitude in a 6$''$ aperture (16.39) is over a magnitude fainter than that 
found in the literature (Table 5). 

\begin{table}
\begin{center}
\begin{tabular}{lll}
\multicolumn{3}{c}{TABLE 5}\\ 
\multicolumn{3}{c}{The range of NIR photometry from the literature.}\\
\hline\hline\\
Name       & \multicolumn{1}{c}{m$_H$ } &  \multicolumn{1}{c}{References}  \\
\hline\\
PKS $0336-019 $& 15.00 - 15.28 & Lepine \etal 1985\\
	   &               & Mead \etal 1990  \\
PKS $0403-132 $& 15.14         & Wright \etal 1983 \\
PKS $0405-123 $& 13.04 - 13.23 & Sun \& Malkan 1989\\ 
	   &               & Wright \etal 1983 \\
PKS $0420-014 $& 12.60 - 15.48 & Gear \etal 1986 \\ 
	   &               & Sitko \& Sitko 1991 \\
PKS $0637-752 $& 13.31         & Hyland \etal 1982 \\
PKS $0736+017 $& 12.18 - 13.95 & Litchfield \etal 1994\\ 
	   &               & Lepine \etal 1985 \\
PKS $1055+018 $& 14.44         & Lepine \etal 1985 \\
PKS $1226+023 $& 10.32 - 10.96 & McAlary \etal 1983\\
	   &               & O'Dell \etal 1978 \\
PKS $1253-055 $& 11.15 - 14.54 & Kidger \etal 1992 \\
	   &               & Roellig \etal 1986  \\
PKS $1510-089 $& 13.12 - 14.09 & Sitko \etal 1983\\ 
           &               & Mead \etal 1990\\
PKS $1954-388 $& 13.72         & Glass 1981 \\
PKS $2128-123 $& 14.06         & Elvis \etal 1994 \\
PKS $2145+067 $& 14.28         & Neugebauer \etal 1979 \\
PKS $2243+123 $& 14.94         & Wright \etal 1983 \\
PKS $2345-167 $& 15.61 - 15.85 & Brindle \etal 1986\\ 
	   &               & Bersanelli \etal 1992 \\
\hline \\ 
\end{tabular}
\end{center}
\end{table}
\noindent 

{\bf PKS 0403--132}. The H band image of this z = 0.571 source is marginally 
resolved, elongated roughly NE--SW, with possible fainter extended emission to
N. Note also several other sources in the field, that may be companions. Our 
H band magnitude (14.97) agrees well with literature photometry (Table 5). 
R\"{o}nnback \etal (1996) studied this source as part of their R band survey 
of intermediate redshift RLQs. They derive M(H,host) $\sim$--25.5 and R(e) = 
25 kpc. Both values are in excellent agreement with our results (see Table 3). 

{\bf PKS 0405--123}. This source at z = 0.574 is well resolved in our image. 
The host galaxy is elongated roughly N--S. Our photometry (H = 13.33) is 
slightly fainter, but in overall agreement with literature photometry (Table 
5). 

{\bf PKS 0420--014}. This high redshift (z = 0.915) source is surprisingly 
well resolved in our image. There is a large range in the magnitudes quoted 
for this object that reflects the rather large flux variability of this object 
(Table 5). 

{\bf PKS 0440--003}. The H band image of this z = 0.844 source remains 
unresolved. No previous NIR photometry was found for PKS 0440--003. 

{\bf PKS 0454--463}. This z = 0.858 source is unresolved in our image. No 
previous NIR photometry was found for PKS 0454--463.

{\bf PKS 0605--085}. The redshift (z = 0.872) was derived from one strong 
broad emission line interpreted as MgII 2800 \AA ~and a rather faint emission 
line of [NeV]3426 \AA ~(Wills 1976). In spite of the high redshift, the H 
band image of PKS 0605--085 appears well resolved. There is, however, a close 
companion object $\sim$3$''$ to SW, likely to be a red M--type galactic star 
(Wills 1976). The extended (possibly dust) emission from this star makes the 
derivation of the luminosity profile of PKS 0605--085 problematic, and we 
prefer only to show the profile, obtained by excluding the sector containing 
the companion, without attempting to model it. No previous NIR photometry was 
found for PKS 0605--085.

{\bf PKS 0637--752}. This z = 0.654 source is marginally resolved in our 
image. There is another object N of it, with possible extension from PKS 
0637--752 toward it, suggesting a physical companion. There are other, bright 
sources toward NE and S, and a jet--like feature to SE. Our H band magnitude 
(14.79) is over a magnitude fainter than that found in the literature (Table 
5). 

{\bf PKS 0736+017}. This nearby well--studied source (z = 0.191) is well 
resolved in our image. The host reveals a rather smooth morphology. Our H band 
photometry (H = 13.66) agrees quite well with most previous studies (Table 5). 
T96 could not discriminate between a bulge or a disk fit for this source, 
although their disk model gave a slightly better fit. Their bulge (disk) fit 
resulted in M(H,host) = --26.1 (--24.9), M(H,nuc) = --27.3 (--27.0), LN/LG = 
3.2 (10.2), and R(e) = 29 (38) kpc. Our result (Table 3) agrees reasonably 
well for the host and nuclear luminosity (both M(H) = --26.2), but we find 
much lower LN/LG ratio (1.0) and much lower effective radius (R(e) = 3.2 kpc). 
In the optical, the source has been imaged with sub--arcsec resolution and 
found to be well fit by a point source and an elliptical galaxy of M(R) = 
--23.5 and R(e) = 15 kpc (Falomo 1996). This yields R--H color of the host of 
$\sim$2.7. 

{\bf PKS 1055+018}. The H band image of this z = 0.888 source remains 
unresolved. Our H band magnitude (15.19) is somewhat fainter than that found 
in the literature (Table 5). 

{\bf PKS 1226+023 = 3C 273}. This extensively studied high--luminosity AGN at 
z = 0.158 is well resolved and shows quite a smooth morphology. Our 
photometry (H = 10.92) agrees well with most previous studies (Table 5). 
Early optical studies of the host galaxy of 3C 273 include Wyckoff \etal 
(1980), Tyson, Baum \& Kreidl (1982) and Hutchings \& Neff (1991), who found 
M(V)$\sim$--22.8, --24.0, and --23.2, respectively, for the host. More 
recently, McLeod \& Rieke (1994a) derived M(H,host) = --26.8, and Bahcall 
\etal (1997) present analysis of the 3C 273 host as part of their sample of 
20 nearby luminous quasars observed with the HST. From a 2--D fit with a 
point source and a bulge, they derive M(V,host) = --23.6. Our result (Table 
3) of M(H,host) = --27.0 is in good agreement with all these previous studies.

{\bf PKS 1253--055 = 3C 279}. The H band image of this well--studied source 
at z = 0.538 was taken under poor seeing conditions and remains unresolved. 
Our H band magnitude (12.92) agrees reasonably well with previous photometry
(Table 5) of this strongly variable quasar. Veron--Cetty \& Woltjer (1990) 
derived M(V) = --25.0 for the host galaxy of 3C 279. For typical galaxy 
colour, this would indicate M(H)$\sim$--28.0, but the host remains unresolved 
in our observations with an upper limit of M(H)$\geq$--29.0. 

{\bf PKS 1504--166}. This source at z = 0.876 remains unresolved in our 
image. The target lies near the border of the array, where the surrounding 
region is quite noisy. No previous NIR photometry was found for PKS 
1504--166. 

{\bf PKS 1510--089}. The H band image of this z = 0.361 source remains 
unresolved. Our H band magnitude (13.29) is much brighter than in most 
previous studies (Table 5). Veron--Cetty \& Woltjer (1990) derived M(V) = 
--22.6 for the host galaxy of PKS 1510--089, indicating for typical galaxy 
colour M(H)$\sim$--25.5 but the host remains unresolved in our data, probably 
due to poor seeing during the observations. An upper limit of M(H) = --26.0 is 
derived. 

{\bf PKS 1954--388}. The H band image of this z = 0.626 source is resolved, 
with a possible companion to NW. Our H band photometry (14.15) agrees 
reasonably well with literature (Table 5). 

{\bf PKS 2128--123}. This source at z = 0.501 is marginally resolved in our 
image. There is a possible companion to NE. Our H band magnitude (14.32) 
agrees well with literature (Table 5). Disney \etal (1995) have studied PKS 
2128--123 with the HST. They found M(R,host) = --23.7, M(R,nuc) = --26.6, 
LN/LG = 13.5 and R(e) = 37.4 kpc. Our result (Table 3) indicates M(H,host) = 
--25.1, LN/LG = 44.0 and R(e) = 16 kpc. While the host magnitude we derive is 
$\sim$1 mag fainter than expected from the I--band for normal galaxy colours, 
this difference is not unreasonable, taking into account all the 
uncertainties in the derivation of the parameters for the marginally resolved 
hosts. 

{\bf PKS 2145+067}. The H band image of this z = 0.990 source is marginally 
resolved. Our H band magnitude (14.47) agrees well with literature (Table 5). 

{\bf PKS 2243--123}. This source at z = 0.630 is marginally resolved in our 
image. There is a possible companion source to SW. Our H band magnitude 
(14.91) is in excellent agreement with literature (Table 5). 

{\bf PKS 2345--167}. The H band image of this source at z = 0.576 is clearly 
resolved. Our H band photometry (13.43) is over 2 magnitudes brighter than 
found in the literature (Table 5).

\end{document}